\documentclass{aa}
\begin{document}
   \title{Images for a Binary Gravitational Lens from a Single Real
   Algebraic Equation} 

   \author{Hideki Asada
%          \inst{1}
          }

   \offprints{H. Asada}

   \institute{Faculty of Science and Technology, 
              Hirosaki University, Hirosaki 036-8561, Japan, 
              \email{asada@phys.hirosaki-u.ac.jp}
             }

   \date{Received ; accepted }

   \abstract{
It is shown that the lens equation for a binary gravitational lens 
being a set of two coupled real fifth-order algebraic equations 
(equivalent to a single complex equation of the same order) can be 
reduced to a single real fifth-order algebraic equation, which 
provides a much simpler way to study lensing by binary objects.  
   \keywords{Gravitational lensing --
                Binaries: general --
                Planetary systems                }
   }

   \maketitle
%
%________________________________________________________________

\section{Introduction}
The gravitational lensing due to a binary system has attracted  
a lot of interest since the pioneering work by Schneider and 
Wei{\ss} (1986). The lens equation adopted until now is a set of 
two coupled real fifth-order equations, equivalently a complex 
fifth-order equation (Witt 1990) which is based on a complex notation 
introduced by Bourassa, Kantowski and Norton (1973, 1975). 

The number of images is classified by curves called caustics, 
on which the Jacobian of the lens mapping vanishes 
on a source plane. 
Caustics for two-point masses are 
investigated in detail and locations of caustics are clarified 
based on a set of two coupled real fifth-order equations 
as an application of catastrophe theory (Erdl and Schneider 1993) 
and on a complex formalism (Witt and Petters 1993), which 
is developed as an efficient method to compute microlensed 
light curves for point sources (Witt 1993): 
In the binary lensing, three images appear for a source 
outside the caustic, while five images are caused for a source 
inside the caustic. 
For a symmetric binary with two equal masses, the lens equation 
for a source on the symmetry axes of the binary becomes 
so simple that we can find the analytic solutions 
(Schneider and Wei{\ss} 1986). 
In star-planet systems, the mass ratio of the binary is so small 
that we can find approximate solutions in general 
(Bozza 1999, Asada 2002). The approximate solutions are used to 
study the shift of the photocenter position by the astrometric 
microlensing (Asada 2002).  

Nevertheless, it is quite difficult to solve these equations, 
since there are no well-established methods for solving coupled 
nonlinear equations numerically with sufficient accuracy 
(Press et al. 1988). 
We show that the lens equation can be reduced to a single 
master equation which is fifth-order in a real variable with 
real coefficients. As a consequence, the new formalism provides 
a much simpler way to study the binary lensing. 

\section{Lens Equation for a Binary System} 
We consider a binary system of two bodies with mass 
$M_1$ and $M_2$ and separation vector $\mbox{\boldmath $L$}$ 
from the object 1 to 2. 
For a later convenience, let us define the Einstein ring radius 
angle as 
\begin{equation}
\theta_{\mbox{E}}=
\sqrt{\frac{4GM D_{\mbox{LS}}}{c^2 D_{\mbox{L}} D_{\mbox{S}}}} ,  
\end{equation}
where $G$ is the gravitational constant, $M$ is the total mass 
$M_1+M_2$, and $D_{\mbox{L}}$, $D_{\mbox{S}}$ and $D_{\mbox{LS}}$ 
denote distances 
between the observer and the lens, 
between the observer and the source, and 
between the lens and the source, respectively. 
We choose the position of the object 1 as the coordinate center. 
In the unit of the Einstein ring radius angle, the lens equation 
reads 
\begin{equation}
\mbox{\boldmath $\beta$}=\mbox{\boldmath $\theta$}
-\Bigl( 
\nu_1 \frac{\mbox{\boldmath $\theta$}}{|\mbox{\boldmath $\theta$}|^2} 
+\nu_2 \frac{\mbox{\boldmath $\theta$}
-\mbox{\boldmath $\ell$}}{|\mbox{\boldmath $\theta$}
-\mbox{\boldmath $\ell$}|^2} 
\Bigr) , 
\label{lenseq}
\end{equation}
where $\mbox{\boldmath $\beta$}$ and $\mbox{\boldmath $\theta$}$ 
denote the vectors for the position of the source and image, 
respectively and we defined the mass ratio and the angular 
separation vector as 
\begin{eqnarray}
\nu_1&=&\frac{M_1}{M_1+M_2} , \\
\nu_2&=&\frac{M_2}{M_1+M_2} , \\
\mbox{\boldmath $\ell$}&=&
\frac{\mbox{\boldmath $L$}}{D_{\mbox{L}}\theta_{\mbox{E}}} . 
\end{eqnarray} 
We have an identity $\nu_1+\nu_2=1$. 
For brevity's sake, $\nu_2$ is denoted by $\nu$. 
Equation (\ref{lenseq}) is a set of two coupled real fifth-order 
equations for $(\theta_x, \theta_y)$, equivalent to a single complex 
fifth-order equation for $\theta_x+i \theta_y$ (e.g. Witt 1990, 1993). 

Let us introduce polar coordinates whose origin is located at 
the mass $M_1$ and the angle is measured from the separation axis 
of the binary. 
The coordinates for the source, image and separation vector are 
denoted by $(\beta_x, \beta_y)=(\rho\cos\varphi, \rho\sin\varphi)$, 
$(\theta_x, \theta_y)=(r\cos\phi, r\sin\phi)$ and 
$(\ell_x, \ell_y)=(\ell, 0)$, respectively, 
where $\rho$, $r$ and $\ell \geq 0$. 
For brevity's sake, $\cos\varphi$ and $\sin\varphi$ are written as 
$C$ and $S$, respectively. The parallel and vertical parts of 
Eq. (\ref{lenseq}) with respect to the separation vector 
$\mbox{\boldmath $\ell$}$ are 
\begin{eqnarray}
&&r\cos\phi\Bigl(1-\frac{1-\nu}{r^2}-
\frac{\nu}{r^2-2\ell r\cos\phi+\ell^2}\Bigr) \nonumber\\
&&=\rho C-\frac{\nu\ell}{r^2-2\ell r\cos\phi+\ell^2} , 
\label{parallel}\\
&&r\sin\phi\Bigl(1-\frac{1-\nu}{r^2}-
\frac{\nu}{r^2-2\ell r\cos\phi+\ell^2}\Bigr)
=\rho S . 
\label{vertical} 
\end{eqnarray}

First, let us investigate a case of $\rho=0$, 
which corresponds to the source being located behind object 1. 
Then, Eq. ($\ref{parallel}$) becomes 
\begin{eqnarray}
&&r\cos\phi\Bigl(1-\frac{1-\nu}{r^2}-
\frac{\nu}{r^2-2\ell r\cos\phi+\ell^2}\Bigr) \nonumber\\
&&=-\frac{\nu\ell}{r^2-2\ell r\cos\phi+\ell^2} , 
\label{parallel2} 
\end{eqnarray}
whose right-hand side does not vanish because of $\nu \neq 0$ for 
the binary. This leads to $\sin\phi=0$, since the R. H. S. 
of Eq. ($\ref{vertical}$) does vanish. 
By using the Cartesian coordinates 
$(\theta_x, \theta_y)=(r\cos\phi, 0)$ and $\theta_x^2=r^2$, 
Eq. ($\ref{parallel2}$) is rewritten as 
\begin{equation}
\theta_x^3-\ell \theta_x^2-\theta_x+\ell(1-\nu)=0 . 
\label{x3}
\end{equation}
All solutions for this equation will be given later 
by Eqs. ($\ref{x}$) and ($\ref{sigma}$) as the limit of $\rho \to 0$. 
In the following, let us assume that $\rho \neq 0$.  
As for $S$, there are two cases, off-axis sources ($S \neq 0$) 
or sources on the symmetry axis ($S=0$).

\subsection{Off-axis sources}
Here, we consider the case that $S$ does not vanish. 
In order to eliminate the common factor in the L. H. S. of 
Eqs. ($\ref{parallel}$) and ($\ref{vertical}$), we divide 
Eq. ($\ref{parallel}$) by Eq. ($\ref{vertical}$), whose R. H. S. 
does not vanish for $S \neq 0$. 
We obtain 
\begin{equation}
r^2-2\ell r\cos\phi+\ell^2=\frac{\nu\ell}
{\rho(C-S\cot\phi)} . 
\label{supplement}
\end{equation}
Substituting this into $r^2-2\ell r\cos\phi+\ell^2$ 
in Eq. ($\ref{vertical}$) gives us 
\begin{equation}
\Bigl[\ell-\rho(C-S\cot\phi)\Bigr]r^2\sin\phi 
-\ell\rho S r-\ell(1-\nu)\sin\phi=0 .  
\end{equation}
We eliminate $r^2$ in this equation by using
Eq. ($\ref{supplement}$). 
Hence, we obtain 
\begin{equation}
r=\frac{R_1(\phi)}{R_2(\phi)} , 
\label{r}
\end{equation}
where we defined 
\begin{eqnarray}
R_1(\phi)&=&
(\ell^2\rho C-\ell\rho^2 C^2-\nu\ell+\rho C)\sin^2\phi
\nonumber\\
&&-\rho S (\ell^2-2\ell\rho C+1)\sin\phi\cos\phi 
\nonumber\\
&&-\ell\rho^2 S^2 \cos^2\phi , \\
R_2(\phi)&=&\rho (C \sin\phi-S \cos\phi) 
\nonumber\\
&&\times\Bigl[(\ell-\rho C)\sin2\phi+\rho S \cos2\phi \Bigr] . 
\end{eqnarray}
We can show that $r \cos\phi$ is a function only 
of $\tan\phi$. 
Namely, Eq. ($\ref{r}$) is rewritten as 
\begin{equation}
r \cos\phi=\frac{\tilde R_1(\tan\phi)}{\tilde R_2(\tan\phi)} , 
\label{rcos}
\end{equation}
where we defined 
\begin{eqnarray}
\tilde R_1(\tan\phi)&=&
(\ell^2\rho C-\ell\rho^2 C^2-\nu\ell+\rho C)\tan^2\phi 
\nonumber\\
&&-\rho S (\ell^2-2\ell\rho C+1)\tan\phi 
-\ell\rho^2 S^2 , \\
\tilde R_2(\tan\phi)&=&\rho (C \tan\phi-S) 
\nonumber\\
&&\times\Bigl[\rho S+2(\ell-\rho C)\tan\phi
-\rho S \tan^2\phi \Bigr] . 
\end{eqnarray}
Equation $(\ref{rcos})$ plays a crucial role in this letter; 
if we find out $\tan\phi$, Eq. ($\ref{rcos}$) gives us 
the value of $r \cos\phi$. The remaining task is deriving 
an equation for $\tan\phi$. 

Let us substitute Eq. ($\ref{r}$) into Eq. ($\ref{supplement}$), 
so that $r$ can be eliminated. 
After lengthy but straightforward computations, we obtain 
an equation for $\tan\phi$ 
\begin{eqnarray}
&&(a_5\tan^5\phi+a_4\tan^4\phi+a_3\tan^3\phi
+a_2\tan^2\phi \nonumber\\
&& +a_1\tan\phi+a_0)\tan\phi = 0 , 
\label{theta}
\end{eqnarray}
where by the frequent use of $C^2+S^2=1$ we defined 
\begin{eqnarray}
a_0&=&\nu\ell\rho^3S^3 , 
\\
a_1&=&\rho^2S^2+2\ell\rho^3CS^2
-\ell^2(2\rho^2S^2-\rho^4S^2)
\nonumber\\
&&-2\ell^3\rho^3CS^2+\ell^4\rho^2S^2
\nonumber\\ 
&&-\nu(5\ell\rho^3CS^2-4\ell^2\rho^2S^2) , 
\\
a_2&=&-2\rho^2CS-4\ell\rho^3(C^2S-S^3) 
\nonumber\\ 
&&+\ell^2(4\rho^2CS-2\rho^4CS)
+4\ell^3\rho^3C^2S-2\ell^4\rho^2CS 
\nonumber\\
&&+\nu\Bigl[\ell(2\rho S+8\rho^3C^2S-2\rho^3S^3) \nonumber\\
&&~~~~~-10\ell^2\rho^2CS+2\ell^3\rho S\Bigr] , 
\\
a_3&=&\rho^2+\ell(2\rho^3C^3-10\rho^3CS^2)
\nonumber\\
&&+\ell^2(-2\rho^2C^2+2\rho^2S^2+\rho^4) 
-2\ell^3\rho^3C+\ell^4\rho^2 \nonumber\\
&&+\nu\Bigl[\ell(-2\rho C-4\rho^3C^3+6\rho^3CS^2) 
\nonumber\\
&&~~~~~+6\ell^2\rho^2C^2-2\ell^3\rho C\Bigr]  \nonumber\\
&&+\nu^2\ell^2 , 
\\
a_4&=&-2\rho^2CS+8\ell\rho^3C^2S-\ell^2(4\rho^2CS+2\rho^4CS)
\nonumber\\
&&+4\ell^3\rho^3C^2S-2\ell^4\rho^2CS 
\nonumber\\
&&+\nu\Bigl[\ell(2\rho S-4\rho^3C^2S+\rho^3S^3)
\nonumber\\
&&~~~~~-2\ell^2\rho^2CS+2\ell^3\rho S\Bigr] , 
\\
a_5&=&\rho^2C^2-2\ell\rho^3C^3+\ell^2(2\rho^2C^2+\rho^4C^2)
\nonumber\\
&&-2\ell^3\rho^3C^3+\ell^4\rho^2C^2 
\nonumber\\
&&+\nu\Bigl[-\ell(2\rho C+\rho^3CS^2) 
+2\ell^2\rho^2C^2-2\ell^3\rho C\Bigr]  
\nonumber\\
&&+\nu^2\ell^2 .  
\end{eqnarray}
It should be noted that all of these coefficients $a_0, \cdots a_5$ 
are not singular, since they are polynomials in $\ell$, $\rho$, 
$\nu$, $C$ and $S$ all of which are finite. 
In the case of nonvanishing $\rho$ and $S$, Eq. ($\ref{vertical}$) 
means that $\sin\phi$ does not vanish and consequently 
neither $\tan\phi$. 
Hence, Eq. ($\ref{theta}$) is reduced to the fifth-order 
equation for $\tan\phi$, 
\begin{equation}
\sum_{i=0}^5a_i(\tan\phi)^i=0 , 
\label{theta2}
\end{equation}
As shown by Galois in the 19th century, a fifth-order equation 
cannot be solved in the algebraic manner 
(e.g. van der Waerden 1966). 
Hence, by solving numerically Eq. $(\ref{theta2})$, 
the image position is obtained 
as $(\theta_x, \theta_y)=(r \cos\phi, r \cos\phi \tan\phi) $. 
It is important to consider a relation of Eq. $(\ref{theta2})$ 
to the treatment of Witt (1993) in which a single complex 
fifth-order algebraic equation for $z=\theta_x+i\theta_y$ is obtained: 
When we use a relation $\tan\phi=-i(z-\bar z)/(z+\bar z)$, 
Eq. $(\ref{theta2})$ can be derived also from the complex equation 
after lengthy manipulations. 

For a source inside the caustic, Eq. $(\ref{theta2})$ 
has five real solutions corresponding to five images, 
while it has three real and two imaginary solutions 
when the source is outside the caustic (Witt 1990, 
Erdl \& Schneider 1993, Witt \& Petters 1993). 
This criteria can be re-stated algebraically by the use of 
the discriminant $D_5$ for the fifth-order equation 
$(\ref{theta2})$, which takes a rather lengthy form in general, 
namely 59 terms (e.g. van der Waerden 1966).
It is worthwhile to mention that all of real solutions 
for Eq. $(\ref{theta2})$ must exist between $-K$ and $K$, 
where $K$ is the larger one between $1$ and 
$|a_0/a_5|+|a_1/a_5|+|a_2/a_5|+|a_3/a_5|+|a_4/a_5|$.  
(For instance, see the section 66 in van der Waerden 1966). 

In numerical computations, it might be difficult to handle extremely 
large $\tau \equiv \tan\phi$. In such a case, we can separate 
$|\tau| \in [0, \infty)$ 
into $|\tau| \in [0, \tau_c)$ and $|\tau| \in [\tau_c, \infty)$. 
We choose $\tau_c$ so that Eq. $(\ref{theta2})$ can be easily 
solved for $|\tau| \in [0, \tau_c)$. 
On the other hand, for $|\tau| \in [\tau_c, \infty)$, instead of 
Eq. $(\ref{theta2})$, we can solve an equation 
for $\cot\phi \in (-1/\tau_c, 1/\tau_c)$ 
\begin{equation}
\sum_{i=0}^5a_i(\cot\phi)^{5-i}=0 . 
\label{theta3}
\end{equation}

\subsection{Sources on the symmetry axis}
Let us consider the case of vanishing $S$ ($C=\pm 1$), namely sources 
on the symmetry axis, for which analytic solutions can be obtained: 
For a binary with two equal masses, explicit solutions were found 
by Schneider and Wei{\ss} (1986), while no explicit solutions 
have been given for an arbitrary mass ratio until now. 
Hence, analytic solutions which are given below are useful 
for verification of numerical implementations, since 
numerical solutions for $S \neq 0$ in the subsection 2.1 
must approach analytic ones as $S \to 0$. 

For $S=0$, Eq. $(\ref{vertical})$ implies apparently the following 
three cases 
\begin{eqnarray}
&&r=0 , \\
&&\sin\phi=0 , \\
&&1-\frac{1-\nu}{r^2}
-\frac{\nu}{r^2-2\ell r\cos\phi+\ell^2}=0 . 
\label{condition3}
\end{eqnarray}
The case of $r=0$ should be discarded, since the left-hand side of 
Eq. ($\ref{vertical}$) diverges. 

Next, let us consider the case of $\sin\phi=0$. 
For this purpose, it is convenient to use the Cartesian coordinates 
$(\theta_x, \theta_y)=(r\cos\phi, 0)$ and 
$(\beta_x, \beta_y)=(\rho C, 0)\equiv(\bar\rho, 0)$ 
for $\bar\rho \in (-\infty, \infty)$. 
By using $r^2=\theta_x^2$, Eq. $(\ref{parallel})$ is rewritten as 
the third-order equation 
\begin{equation}
\theta_x^3-(\ell+\bar\rho)\theta_x^2
+(\ell\bar\rho-1)\theta_x+\ell(1-\nu)=0 , 
\label{x33}
\end{equation}
which coincides with Eq. ($\ref{x3}$) as $\bar\rho \to 0$. 
Equation ($\ref{x33}$) is solved explicitly as 
\begin{equation}
\theta_x=2\sqrt{-p}\cos\sigma+\frac{\ell+\bar\rho}{3} , 
\label{x}
\end{equation}
with 
\begin{equation}
\sigma=\left\{
\begin{array}{l}
\frac13\cos^{-1}\Bigl(\frac{q}{2p\sqrt{-p}}\Bigr) , \\
\frac13\cos^{-1}\Bigl(\frac{q}{2p\sqrt{-p}}\Bigr)
+\frac{2\pi}{3} , \\
\frac13\cos^{-1}\Bigl(\frac{q}{2p\sqrt{-p}}\Bigr)
+\frac{4\pi}{3} ,  
\label{sigma}
\end{array}\right.
\end{equation}
where 
\begin{eqnarray}
&&p=-\frac19(\ell+\bar\rho)^2+\frac13(\ell\bar\rho-1) , \\
&&q=-\frac{2}{27}(\ell+\bar\rho)^3+\frac13(\ell+\bar\rho)
(\ell\bar\rho-1)+\ell(1-\nu) . 
\end{eqnarray}
Actually, we can show that $p < 0$ and $q^2+4 p^3 < 0$, which mean 
these three solutions exist for any source position.  
They are on-axis solutions, while the remaining two 
solutions are off-axis solutions, which are present only for sources 
within the caustics. 

Finally, for the case defined by Eq. $(\ref{condition3})$, 
Eq. $(\ref{parallel})$ leads to 
\begin{equation}
\frac{\nu}{r^2-2\ell r\cos\phi+\ell^2}
=\frac{\bar\rho}{\ell} .
\label{condition3b}
\end{equation}
Replacing $r^2-2\ell r\cos\phi+\ell^2$ in 
Eq. $(\ref{condition3})$ by this equation, we obtain 
\begin{equation}
r^2=\frac{\ell(1-\nu)}{\ell-\bar\rho} , 
\end{equation}
which has the positive solution  
\begin{equation}
r=\sqrt{\frac{\ell(1-\nu)}{\ell-\bar\rho}} , 
\label{r0}
\end{equation}
if and only if $\ell > \bar\rho$. Substitution of the solution into 
Eq. $(\ref{condition3b})$ leads to 
\begin{equation}
\cos\phi=\frac12\sqrt{\frac{\ell-\bar\rho}{\ell(1-\nu)}}
\Bigl(\ell+\frac{\bar\rho-\nu\ell}{\bar\rho(\ell-\bar\rho)}\Bigr) . 
\label{cos0}
\end{equation}
A condition $|\cos\phi| \leq 1$ is rewritten as 
\begin{eqnarray}
&&\ell^2\bar\rho^4-2\ell(\ell^2-1+2\nu)\bar\rho^3
+\Bigl[(1-\ell^2)^2+6\nu\ell^2\Bigr]\bar\rho^2 \nonumber\\
&&-2\ell(1+\ell^2)\nu\bar\rho+\nu^2\ell^2 \leq 0 . 
\end{eqnarray}
Only if this is satisfied for $\bar\rho < \ell$, 
the two images appear at the location given by Eqs. ($\ref{r0}$) 
and ($\ref{cos0}$).

\section{Conclusion}
We have carefully reexamined the lens equation for a binary system 
in the polar coordinates. 
As a consequence, we have derived Eq. ($\ref{theta2}$) 
for $\tan\phi$. After solving the equation, $r \cos\phi$ is 
determined by Eq. ($\ref{rcos}$). Hence, the image position 
$(\theta_x, \theta_y)=(r \cos\phi, r \cos\phi \tan\phi) $ 
can be determined. 
Our formulation based on the one-dimensional equation 
($\ref{theta2}$) is significantly useful compared with previous 
two-dimensional treatments for which there are no well-established 
numerical methods (Press et al. 1988); the new formulation enables 
us to study the binary lensing more precisely with saving time 
and computer resources. 
For instance, it is effective in rapid and accurate light-curve 
fitting to microlensing events, in particular 
due to star-planet systems.

\begin{acknowledgements}
The author would like to thank M. Bartelmann, M. Kasai and 
H. Nakazato for useful conversation. 
This work was supported in part by a Japanese Grant-in-Aid 
for Scientific Research from the Ministry of Education, 
No. 13740137 and the Sumitomo Foundation. 
\end{acknowledgements}

\end{document}